\documentclass[11pt,twoside]{article}
\usepackage{asp2010}
\resetcounters

\begin{document}

\title{The flux calibration of Gaia}
\author{Elena~Pancino$^1$
\affil{$^1$INAF --- Osservatorio Astronomico di Bologna, Via C. Ranzani 1,
I-40127 Bologna, Italy}}

\begin{abstract}

The Gaia mission is described, along with its scientific potential and its
updated science perfomancesAlthough it is often described as a self-calibrated
mission, Gaia still needs to tie part of its measurements to external scales (or
to convert them in physical units). A detailed decription of the Gaia
spectro-photometric standard stars survey is provided, along with a short
description of the Gaia calibration model. The model requires a grid of
approximately 200 stars, calibrated to a few percent with respect to Vega, and
covering different spectral types.

\end{abstract}

\section{The Gaia mission}

Gaia is a cornerstone mission of the ESA Space Program, presently scheduled for
launch in 2013. The Gaia satellite will perform an all-sky survey to obtain
parallaxes and proper motions to $\mu as$ precision for about 10$^9$ point-like
sources and astrophysical parameters ($T_{\mathrm{eff}}$, $\log g$, $E(B-V)$,
metallicity etc.) for stars down to a limiting magnitude of $V\simeq 20$, plus
2-30 km/s  accuracy (depending on spectral type), radial velocities for several
millions of stars down to $V < 17$. 

Such an observational effort has been compared to the mapping of the human genome
for the amount of collected data and for the impact that it will have on all
branches of astronomy and astrophysics. The expected end-of-mission astrometric
accuracies are almost 100 times better than the Hipparrcos dataset (see Perryman
et al. 1997). This exquisite precision will allow a full and detailed
reconstruction of the 3D spatial structure and 3D velocity field of the Milky Way
galaxy within $\simeq 10$ kpc from the Sun. This will provide answers to
long-standing questions about the origin and evolution of our Galaxy, from a
quantitative census of its stellar populations, to a detailed characterization of
its substructures \citep[as, for instance, tidal streams in the Halo,
see][]{ibata07}, to the distribution of dark matter. 

The accurate 3D motion of more distant Galactic satellites (as globular clusters
and the Magellanic Clouds) will be also obtained by averaging the proper motions
of many thousands of member stars: this will provide an unprecedented leverage to
constrain the mass distribution of the Galaxy and/or non-standard theories of
gravitation. Gaia will determine direct geometric distances to essentially any
kind of standard candle currently used for distance determination, setting the
whole cosmological distance scale on extremely firm bases. 

As challenging as it is, the processing and analysis of the huge data-flow
incoming from Gaia is the subject of thorough study and preparatory work by the
Data Processing and Analysis Consortium (DPAC), in charge of all aspects of the
Gaia data reduction. The consortium comprises more than 400 scientists from 25
European institutes. Gaia is usually described as a self-calibrating mission, but
it also needs {\em external} data to fix the zero-point of the magnitude system
and radial velocities, and to calibrate the classification/parametrization
algorithms. All these additional data are termed auxiliary data and have to be
available, at least in part, three months before launch. While part of the
auxiliary data already exist and must only be compiled from archives, this is not
true for several components. To this aim a coordinated programme of ground-based
observations is being organized by a dedicated inter-CU committee (GBOG), that
promotes sinergies and avoids duplications of effort.

\begin{figure}
\includegraphics[width=6.8cm,height=3.8cm]{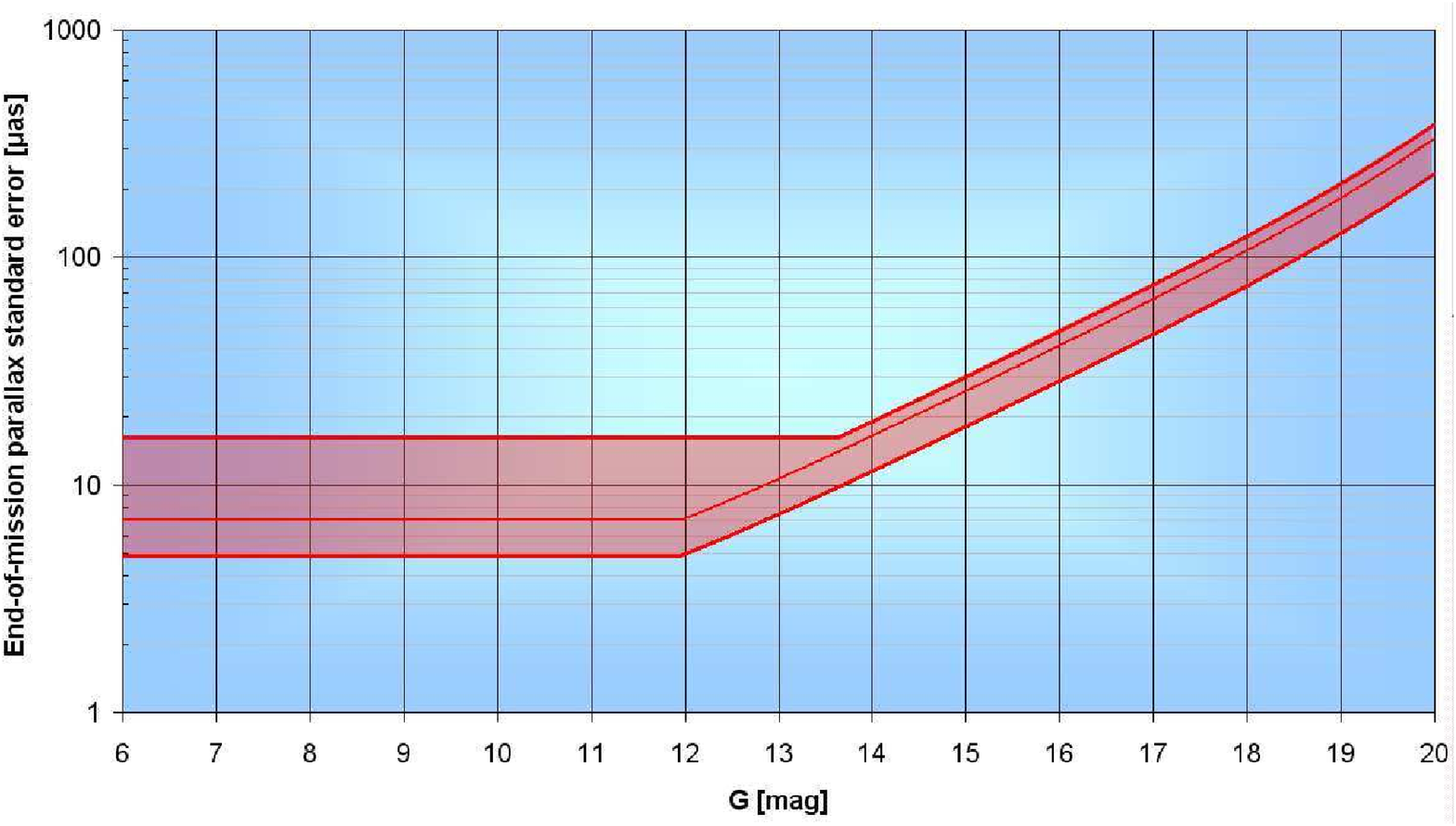}\includegraphics[width=6.8cm,height=3.8cm]{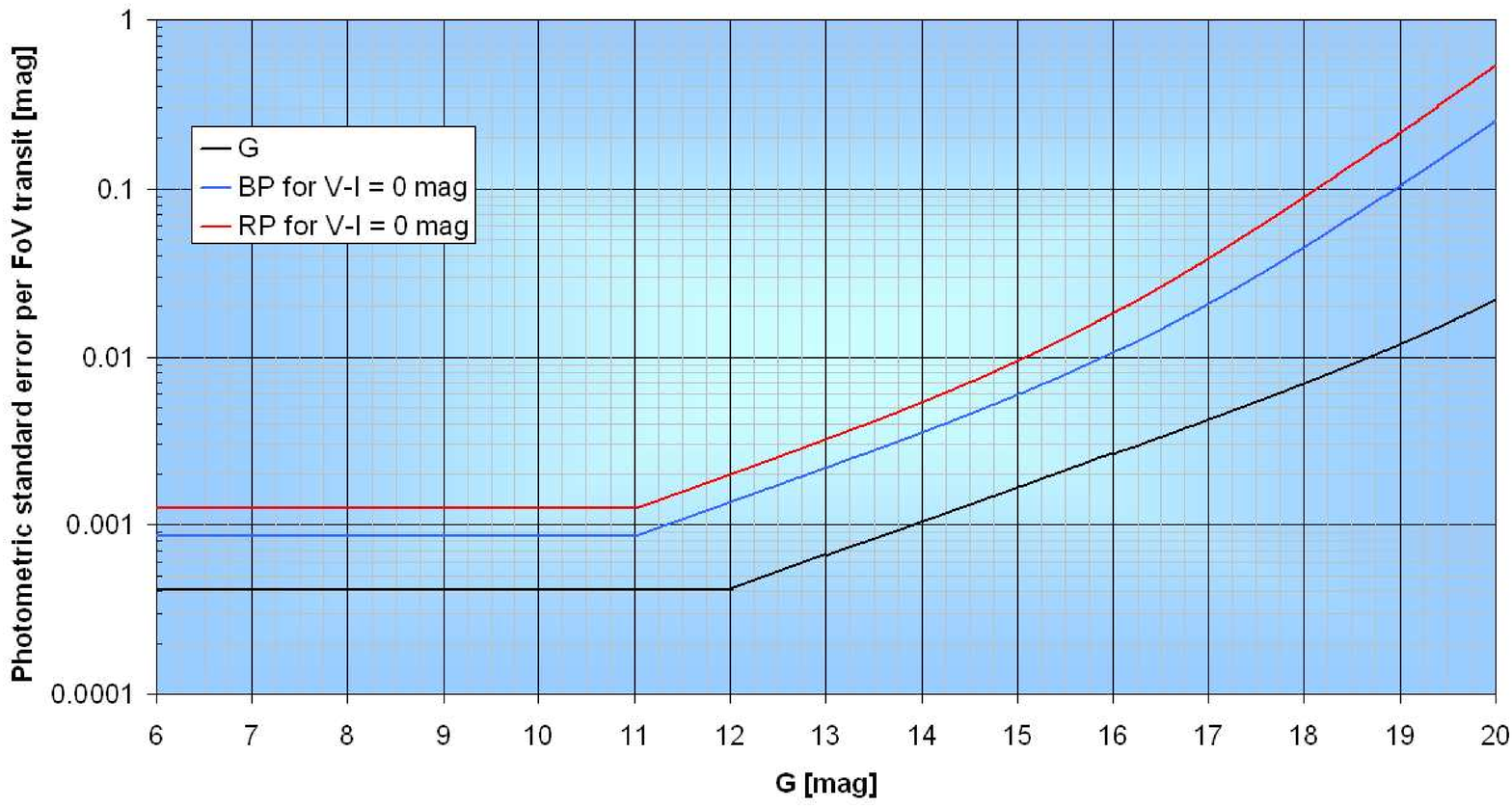}
\includegraphics[width=7cm,height=3.8cm]{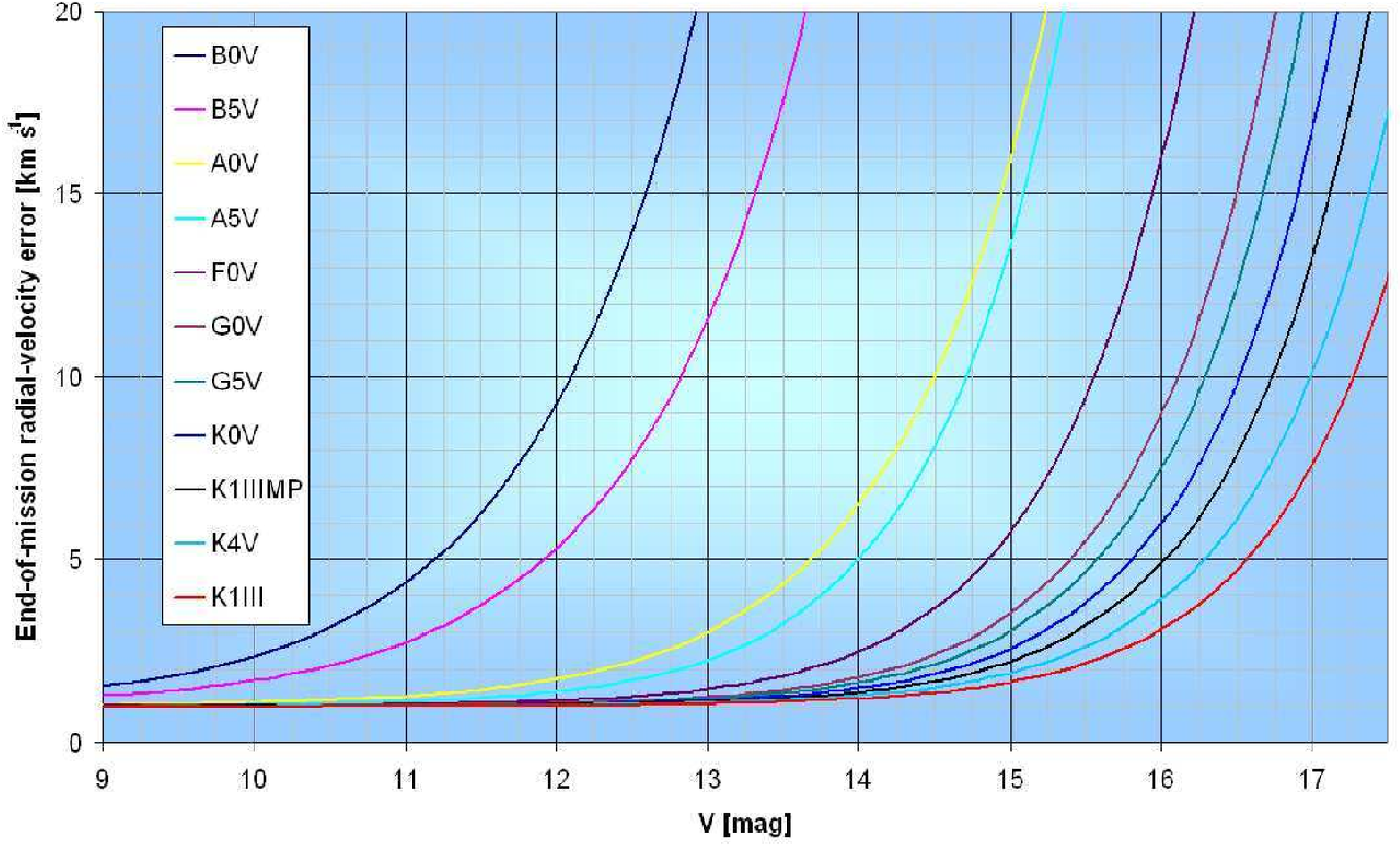}\includegraphics[width=6.5cm,height=3.8cm]{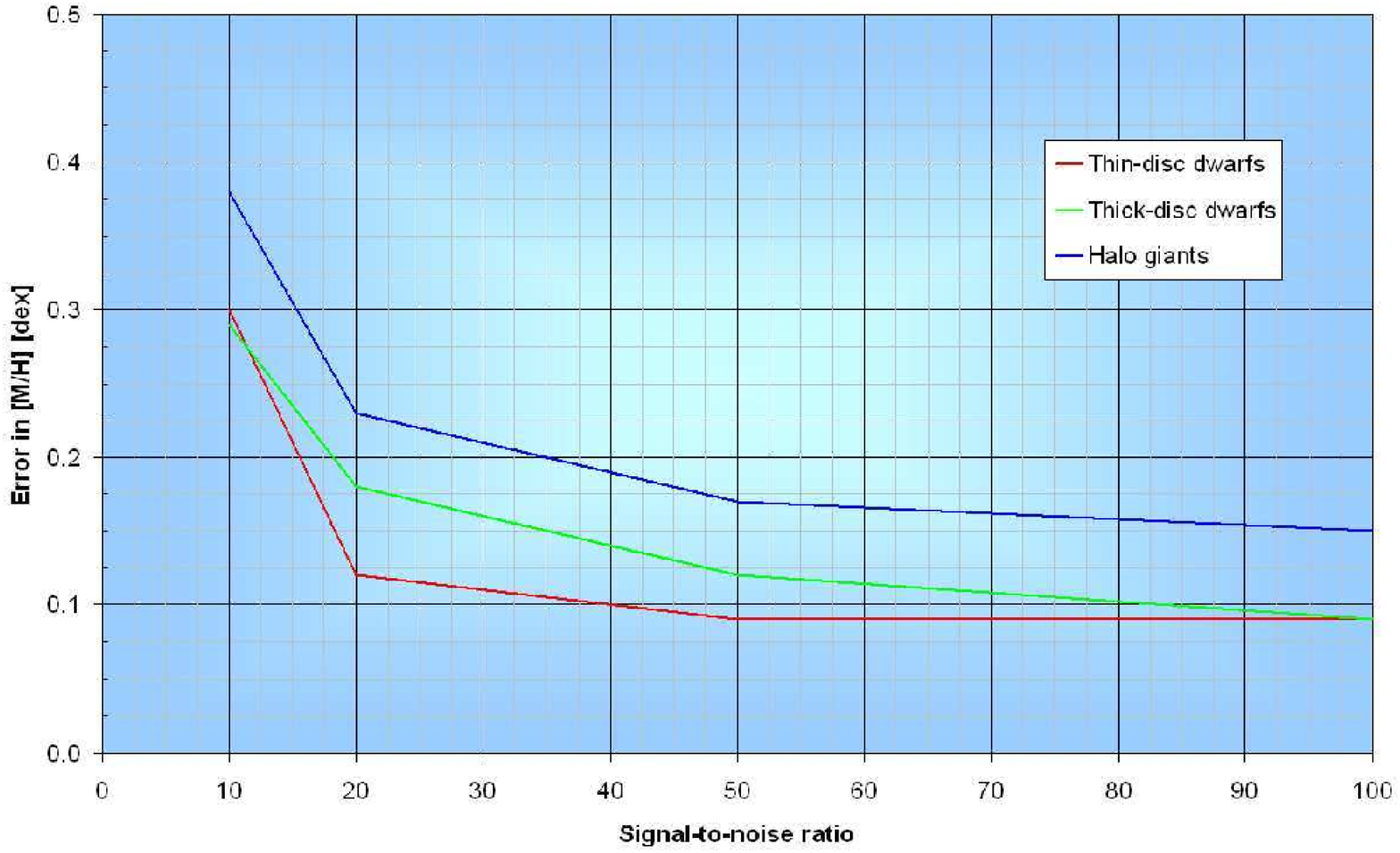}
\caption{The Gaia updated science performances. {\em Top left panel:} The
end-of-mission uncertainty on astrometric measurements (in $\mu$as) as a function
of the integrated G magnitude of a G2V star. {\em Top right panel:}
End-of-mission uncertainties on the G, G$_{\rm{BP}}$, G$_{\rm{RP}}$ magnitudes as
a function of G, for a V--I=0~mag star. {\em Bottom left panel:} End-of-mission
uncertainty on radial velocity (in km/s) as a function of G magnitude for
different spectral types. {\em Bottom right panel:} Uncertainty on the
metallicity ([M/H] in dex) derived from RVS spectra as a function of the S/N
ratio for different Galactic populations. \copyright ESA}
\label{pancino_fig_perf}
\end{figure}

\subsection{Science goals and capabilities}

Gaia will measure the positions, distances, space motions, and many physical
characteristics of some billion stars in our Galaxy and beyond. For many years,
the state of the art in celestial cartography has been the Schmidt surveys of
Palomar and ESO, and their digitized counterparts. The measurement precision,
reaching a few  millionths of a second of arc, will be unprecedented. The most
updated science performances can be found on the Gaia ESA
webpage\footnote{{\url{http://www.rssd.esa.int/index.php?project=GAIA&page=Science_Performance}}},
along with some formulaes to re-compute figures easily, and with all the
references (see Fig.~\ref{pancino_fig_perf}). Some millions of stars will be
measured with a distance accuracy of better than 1 per cent; some 100 million or
more to better than 10 per cent.  Gaia's resulting scientific harvest is of almost
inconceivable extent and implication. 

Gaia will provide detailed  information on stellar evolution and star formation in
our Galaxy. It will clarify the origin and formation history  of our Galaxy. The
results will precisely identify relics of tidally-disrupted accretion debris,
probe the distribution of dark matter, establish the luminosity function for
pre-main sequence stars, detect and categorize rapid evolutionary stellar phases,
place unprecedented constraints on the age, internal structure and evolution of
all stellar types, establish a rigorous distance scale framework throughout the
Galaxy and beyond, and classify star formation and kinematical and dynamical
behaviour within the Local Group of galaxies. 

Gaia will pinpoint exotic objects in colossal and almost unimaginable numbers:
many thousands of extra-solar  planets will be discovered (from both their
astrometric wobble and from photometric transits) and their detailed orbits and
masses determined; tens of thousands of brown dwarfs and white dwarfs will be
identified; tens of  thousands of extragalactic supernovae will be discovered;
Solar System studies will receive a massive impetus through the observation of
hundreds of thousands of minor planets; near-Earth objects, inner Trojans and
even  new trans-Neptunian objects, including Plutinos, may be discovered. 

Gaia will follow the bending of star light by the Sun and major planets over the
entire celestial sphere, and  therefore directly observe the structure of
space-time -- the accuracy of its measurement of General Relativistic light
bending may reveal the long-sought scalar correction to its tensor form. The PPN
parameters $\gamma$ and $\beta$, and the solar quadrupole moment J2, will be
determined with unprecedented precision. All this, and more, through the accurate
measurement of star positions. 

For more information on the Gaia mission: http://www.rssd.esa.int/Gaia. More
information for the public on Gaia and its science capabilities are contained in
the {\em Gaia information
sheets}\footnote{\url{http://www.rssd.esa.int/index.php?pro
ject=GAIA\&page=Info\_sheets\_overview}.}. Excellent reviews of the science
possibilities opened by Gaia can be found in \citet{perryman97} and
\citet{mignard05}.

\subsection{Launch, timeline and data releases}

The first idea for Gaia began circulating in the early 1990, culminating in a
proposal for a  cornerstone mission within ESA's science programme submitted in
1993, and a workshop in Cambridge in June 1995. By the time the final catalogue
will be released approximately in 2020, almost two decades of work will have
elapsed between the orginal concept and mission completion.  

Gaia will be launched by a Soyuz carrier (rather than the initially foreseen
Ariane 5) in 2013 from French Guyana and will start operating once it will reach
its Lissajus orbit around L2 (the unstable Langrange point of the Sun and
Eart-Moon system), in about one month. Two ground stations will receive the
compressed Gaia data during the 5 years\footnote{If -- after careful evaluation --
the scientific output of the mission will benefit from an extension of the
operation period, the satellite should be able to gather data for one more year,
remaining within the Earth eclipse.} of operation: Cebreros (Spain) and Perth
(Australia). The data will then be transmitted to the main data centers throughout
Europe to allow for data processing. We are presently in technical development
phase C/D, and the hardware is being built, tested and assembled. Software
development started in 2006 and is presently producing and testing pipelines with
the aim of delivering to the astrophysical community a full catalogue and dataset
ready for scientific investigation.

Apart from the end-of-mission data release, foreseen around 2020, some
intermediate data releases are foreseen. In particular, there should be one first
intermediate release covering either the first 6 months or the first year of
operation, followed by a second and possibly a third intermediate release, that
are presently being discussed. The data analysis will proceed in parallel with
observations, the major pipelines re-processing all the data every 6 months, with
secondary cycle pipelines -- dedicated to specific tasks -- operating on different
timescales. In particular, verified science alerts, based on unexpected
variability in flux and/or radial velocity, are expected to be released within 24
hours from detection, after an initial period of testing and fine-tuning of the
detection algorithms. 

\subsection{Mission concepts}

\begin{figure}
\plottwo{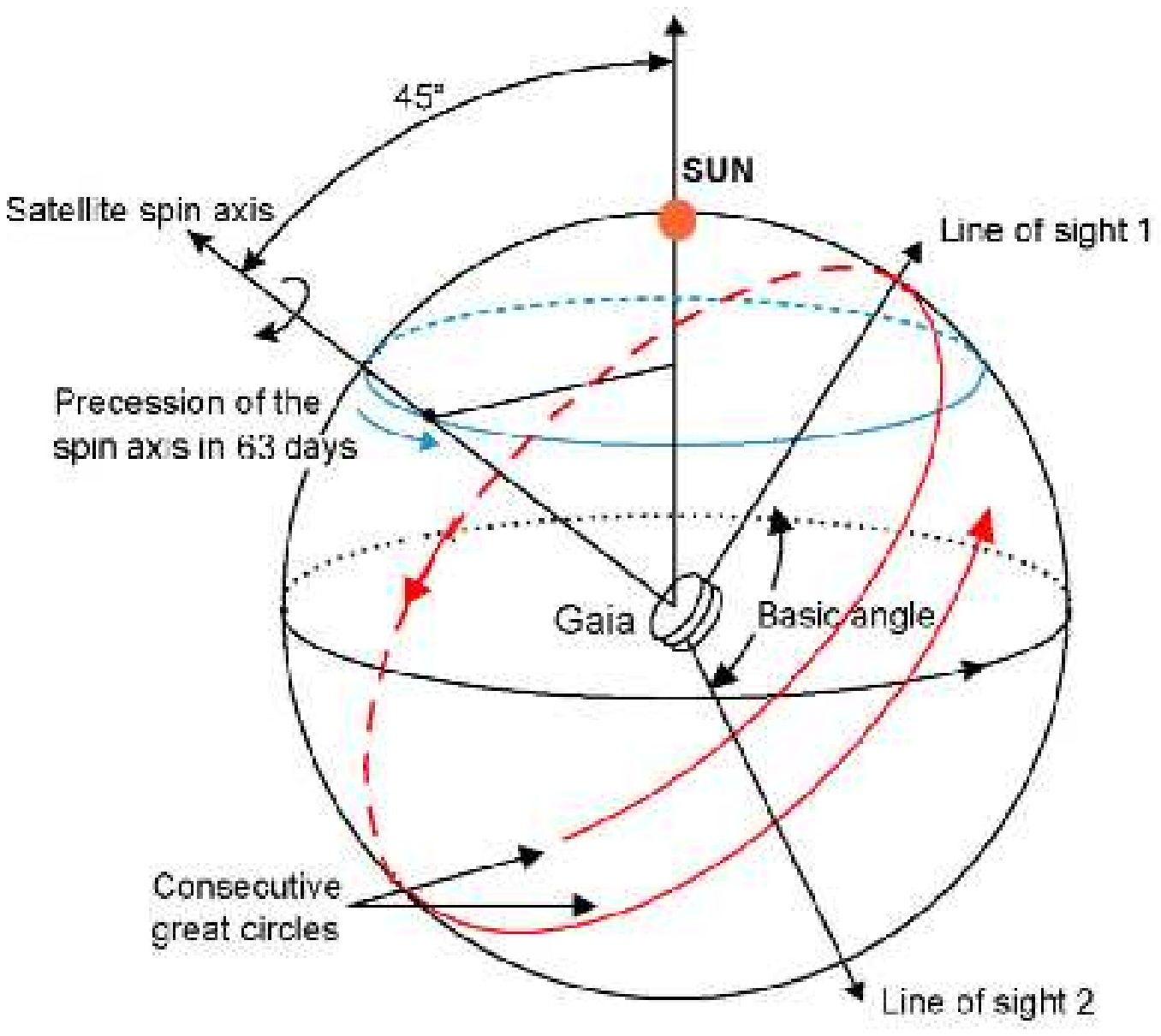}{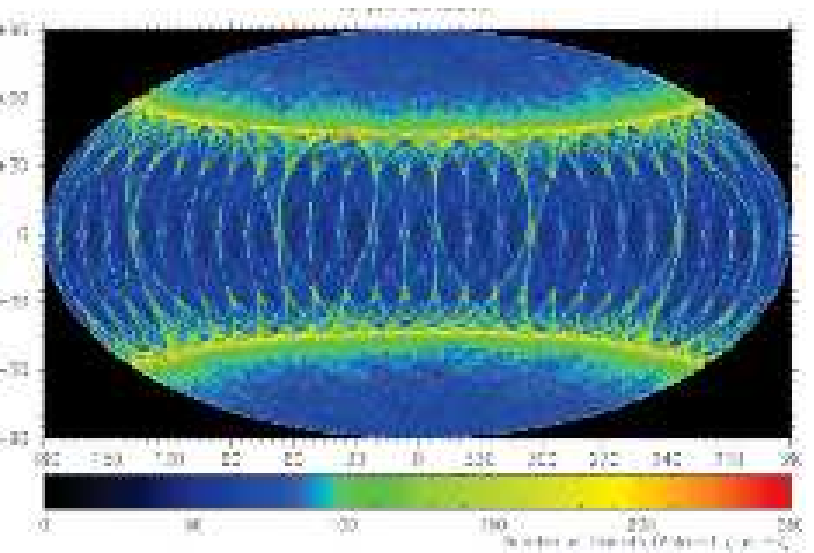}
\caption{Left: the scanning law of Gaia during main operations; Right: the average
number of passages on the sky, in ecliptic coordinates. \copyright ESA}
\label{pancino_fig_scan}
\end{figure}

During its 5-year operational lifetime, the satellite will continuously spin
around its axis, with a constant speed of 60~arcsec/sec. As a result, over a
period of 6 hours, the two astrometric fields of view will scan across all
objects located along the great circle perpendicular to the spin axis
(Figure~\ref{pancino_fig_scan}). As a result of the basic angle of 106.5$^{\rm
o}$ separating the astrometric fields of view on the sky, objects transit the
second field of view with a delay of 106.5 minutes compared to the first field.
Gaia's spin axis does not point to a fixed direction in space, but is carefully
controlled so as to precess slowly on the sky. As a result, the great circle that
is mapped by the two fields of view every 6 hours changes slowly with time,
allowing reapeated full sky coverage over the mission lifetime. The best
strategy, dictated by thermal stability and power requirements, is to let the
spin axis precess (with a period of 63 days) around the solar direction with a
fixed angle of 45$^{\rm o}$. The above scanning strategy, referred to as
``revolving scanning", was successfully adopted during the Hipparcos mission. 

Every sky region will be scanned on average 70-80 times, with regions lying at
$\pm$45$^{\rm o}$ from the Ecliptic Poles being scanned on average more often than
other locations. Each of the Gaia targets will be therefore scanned (within
differently inclined great circles) from a minimum of approximately 10 times to a
maximum of 250 times (Figure~\ref{pancino_fig_scan}, right panel). Only point-like
sources will be observed, and in some regions of the sky, like the Baade's window,
$\omega$ Centauri or other globular clusters, the star density of the two combined
fields of view will be of the order of 750\,000 or more per square degree,
exceeding the storage capability of the onboard processors, so Gaia will not study
in detail these dense areas. 

\subsection{Focal plane}

\begin{figure}
\plotone{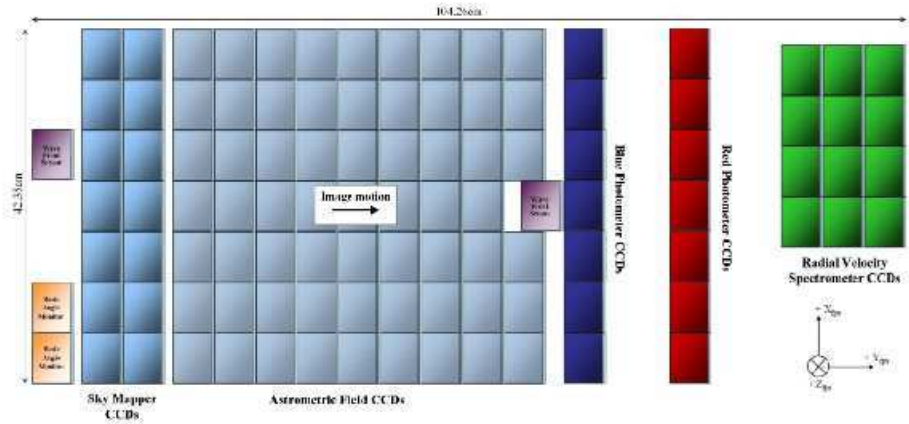}
\caption{The 105 on the Gaia focal plane. \copyright ESA}
\label{pancino_fig_foc}
\end{figure}

Figure~\ref{pancino_fig_foc} shows the focal plane of Gaia, with its 105 CCDs,
which are read in TDI (Time Delay Integration) mode: objects enter the focal
plane from the left and cross one CCD in 4 seconds. Apart from some technical
CCDs that are of little interest in this context, the first two CCD columns, the
Sky Mappers (SM), perform the on-board detection of point-like sources, each of
the two columns being able to see only one of the two lines of sight. After the
objects are identified and selected, small windows are assigned, which follow
them in the astrometric field (AF) CCDs where white light (or G-band) images are
obtained (Section~\ref{pancino-sec-af}). Immediately following the AF, two
additional columns of CCDs gather light from two slitless prism spectrographs,
the blue spectrophotometer (BP) and the red one (RP), which produce dispersed
images (Section~\ref{pancino-sec-phot}). Finally, objects transit on the Radial
Velocity Spectrometer (RVS) CCDs to produce higher resolution spectra around the
Calcium Triplet (CaT) region (Section~\ref{pancino-sec-rvs}). 

\subsection{Astrometry} 
\label{pancino-sec-af}

\begin{figure}
\plottwo{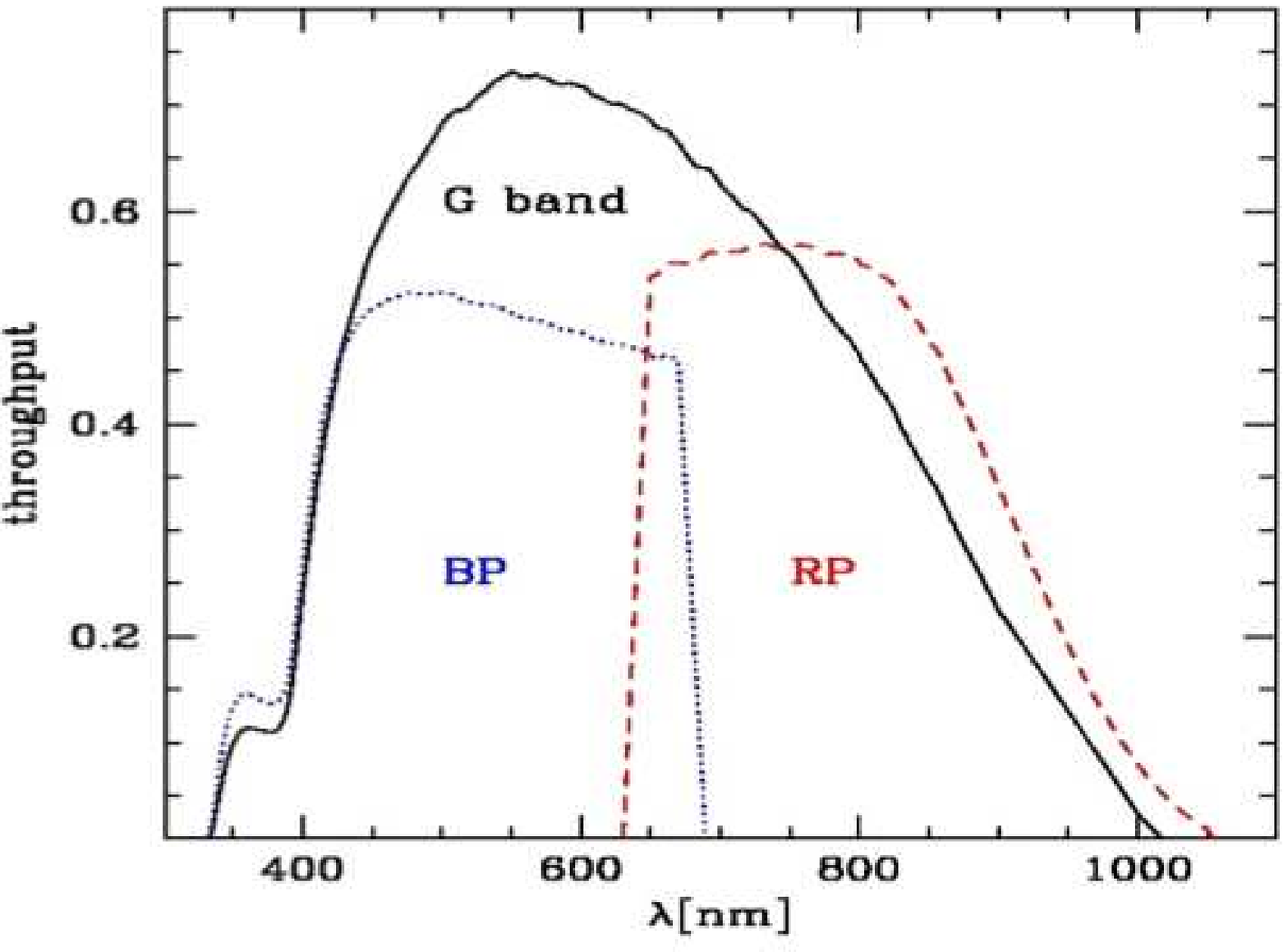}{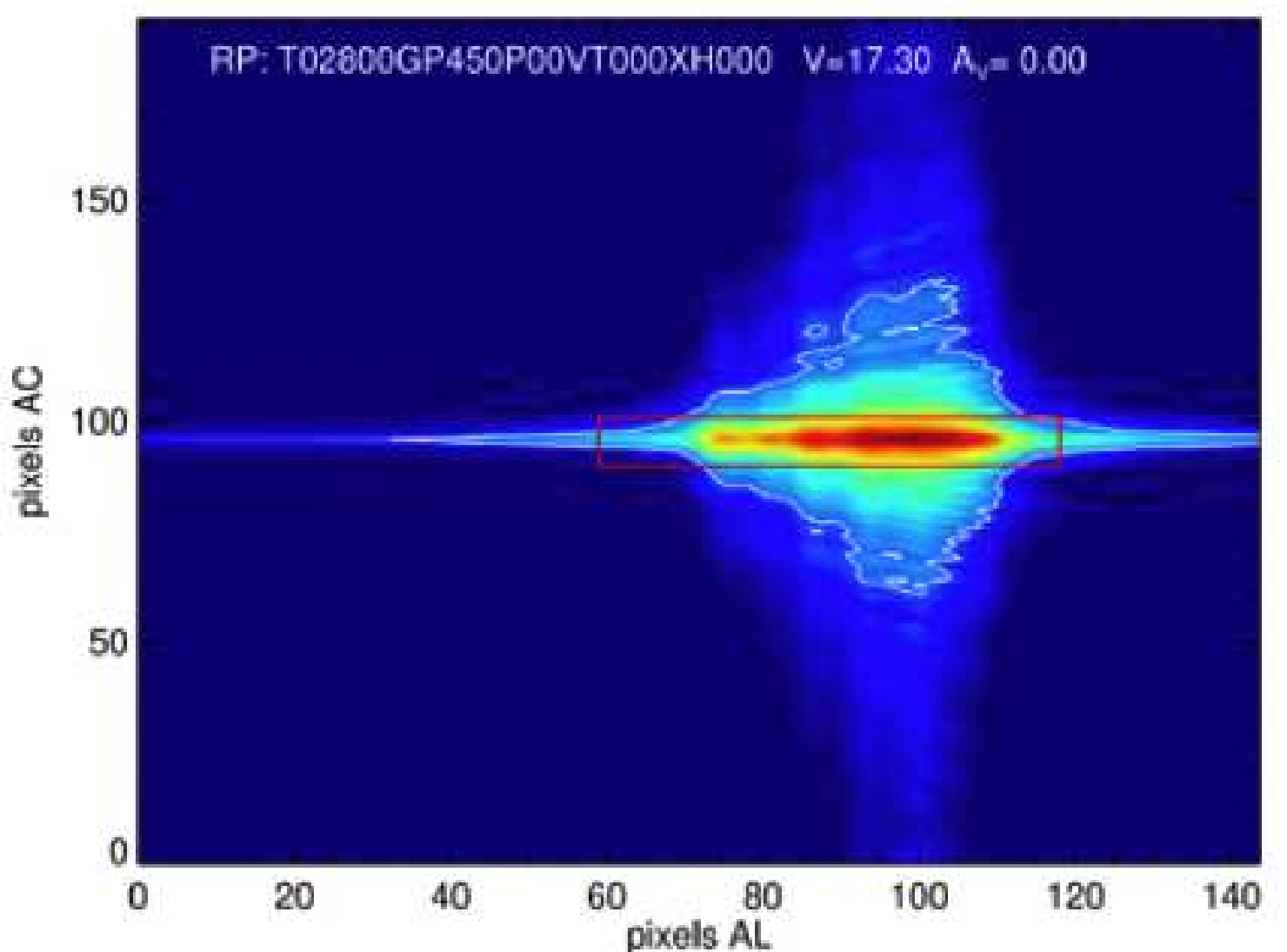}
\caption{Left: the passbands of the G-band, BP and RP; Right: a simulated RP
dispersed image, with a red rectangle marking the window assigned for compression
and ground telemetry. \copyright ESA} \label{pancino_fig_phot}
\end{figure}

The AF CCDs will provide G-band images, i.e., white light images where the
passband is defined by the telescope optics transmission and the CCDs
sensitivity, with a very broad combined passband ranging from 330 to 1050~nm and
peaking around 500--600~nm (Figure~\ref{pancino_fig_phot}). The objective of
Gaia's astrometric data reduction system is the construction of core mission
products: the five standard astrometric parameters, position ($\alpha$,
$\delta$), parallax ($\varpi$), and proper motion ($\mu_{\alpha}$,
$\mu_{\delta}$) for all observed stellar objects. The expected end-of-mission
precision is shown in Figure~\ref{pancino_fig_perf} and discussed in detail in
the Gaia ESA webpages.

To reach these end-of-mission precisions, the average 70--80 observations per
target gathered during the 5-year mission duration will have to be combined in a
self-consistent manner. 40~Gb of telemetry data will first pass through the
Initial Data Treatment (IDT) which determines the image parameters and centroids,
and then performes an object cross-matching. The output forms the so-called One
Day Astrometric Solution (ODAS), together with the satellite attitude and
calibration, to the sub-milliarcsecond accuracy.  The data are then written to
the Main Database. 

The next step is the Astrometric Global Iterative Solution (AGIS) processing.
AGIS processes together the attitude and calibration parameters with  the source
parameters, refining them in an iterative procedure that stops when the
adjustments become sufficiently small. As soon as new data come in, on the basis
of 6 months cycles, all the data in hand are reprocessed toghether from scratch.
This is the only scheme that allows for the quoted precisions, and it is also the
philosophy that justifies Gaia as a self-calibrating mission. The primary AGIS
cycle will treat only stars that are flagged as single and non-variable (expected
to be around 500 millions), while other kinds of objects will be computed in
secondary AGIS cycles that utilize the main AGIS solution. Dedicated pipelines
for specific kinds of objects (asteroids, slightly extended objects, variable
objects and so on) are being put in place to extract the best possible precision.
Owing to the large data volume (100~Tb) that Gaia will produce, and to the
iterative nature of the processing, the computing challenges are formidable: AGIS
processing alone requires some 10$^{21}$~FLOPs which translates to runtimes of
months on the ESAC computers in Madrid.

\subsection{Spectrophotomety}
\label{pancino-sec-phot}

The primary aim of the photometric instrument is mission critical in two respects:
(i) to correct the measured centroids position in the AF for systematic chromatic
effects, and (ii) to classify and determine astrophysical characteristics of all
objects, such as temperature, gravity, mass, age and chemical composition (in the
case of stars).

The BP and RP spectrophotometers are based on a dispersive-prism approach such
that the incoming light is not focussed in a PSF-like spot, but dispersed along
the scan direction in a low-resolution spectrum. The BP operates between
330--680~nm while the RP between 640-1000~nm (Figure~\ref{pancino_fig_phot}). Both
prisms have appropriate broad-band filters to block unwanted light. The two
dedicated CCD stripes cover the full height of the AF and, therefore, all objects
that are imaged in the AF are also imaged in the BP and RP. 

The resolution is a function of wavelength, ranging from 4 to 32 nm/pix for BP and
7 to 15 nm/pix for RP. The spectral resolution, R=$\lambda/\delta \lambda$ ranges
from 20 to 100 approximately. The dispersers have been designed in such a way that
BP and RP spectra are of similar sizes (45 pixels). Window extensions meant to
measure the sky background are implemented. To compress the amount of data
transmitted to the ground, all the BP and RP spectra -- except for the brightest
stars -- are binned on chip in the across-scan direction, and are transmitted to
the ground as one-dimensional spectra. Figure~\ref{pancino_fig_phot} shows a
simulated RP spectrum, unbinned, before windowing, compression, and telemetry.

The final data products will be the end-of-mission (or intermediate releases) of
global, combined BP and RP spectra and integrated magnitudes G$_{\rm{BP}}$ and
G$_{\rm{RP}}$. Epoch spectra will be released only for specific classes of objects,
such as variable stars and quasars, for example. The internal flux calibration of
integrated magnitudes, including the G magnitudes as well, is expected at a
precision of 0.003~mag for G=13 stars, and for G=20 stars goes down to 0.07~mag in
G, 0.3~mag in G$_{\rm{BP}}$ and G$_{\rm{RP}}$. The external calibration should be
performed with a precision of the order of a few percent (with respect to Vega,
see section~\ref{pancino-sec-survey}).

\begin{figure}
\centering
\includegraphics[width=10cm]{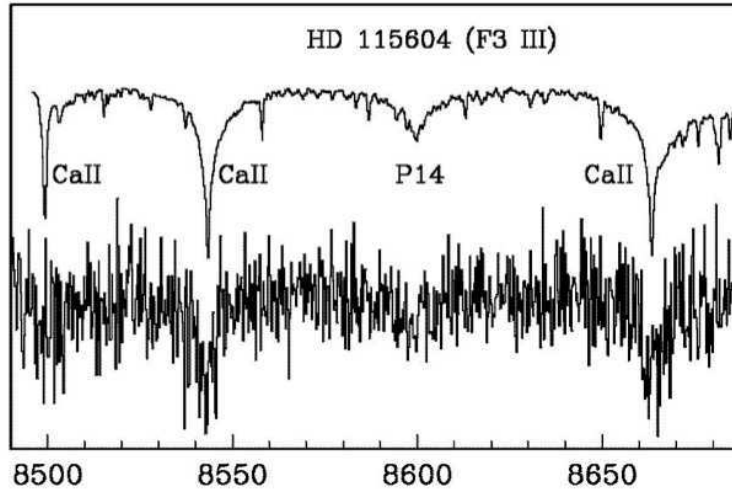}
\caption{Simulated RVS end-of-mission spectra for the extreme cases of 1 single 
transit (bottom spectrum) and of 350 transits (top spectrum). \copyright ESA}
\label{pancino_fig_rvs}
\end{figure}

\subsection{High-resolution spectroscopy}
\label{pancino-sec-rvs}

The primary objective of the RVS is the acquisition of radial velocities, which
combined with positions, proper motions, and parallaxes will provide the means to
decipher the kinematical state and dynamical history of our Galaxy.

The RVS will provide the radial velocities of about 100--150 million stars up to
17-th magnitude with the precisions illustrated in Figure~\ref{pancino_fig_perf}
and in the Gaia ESA webpage. The spectral resolution, R=$\lambda$/$\delta\lambda$
will be 11\,500. Radial velocities will be obtained by cross-correlating observed
spectra with either a template or a mask. An initial estimate of the source
atmospheric parameters will be used to  select  the  most appropriate template or
mask. On average, 40 transits will be collected for each object during the 5-year
lifetime of Gaia, since the RVS does not cover the whole width of the Gaia AF
(Figure~\ref{pancino_fig_foc}). In total, we expect to obtain some 5 billion
spectra (single transit) for the brightest stars. The analysis of this huge
dataset will be complicated, not only because of the sheer data volume, but also
because the spectroscopic data analysis relies on the multi-epoch astrometric and
photometric data. 

The covered wavelength range (847-874 nm, Figure~\ref{pancino_fig_rvs}) is a rich
domain, centered on the infrared calcium triplet: it  will not only provide
radial velocities, but also many stellar and interstellar diagnostics. It has
been selected to coincide with the energy distribution peaks of G anf K type
stars, which are the most abundant targets. In early type stars, RVS spectra may
contain also weak He lines and N, although they will be dominated by the Paschen
lines. The RVS data will effectively complement the astrometric and photometric
observations, improving object classification. For stellar objects, it will
provide atmospheric parameters such as effective temperature, surface gravity,
and individual abundances of key elements such as Fe, Ca, Mg, Si for millions of
stars down to G$\simeq$12. Also, Diffuse Intertellar Bands (DIB) around 862 nm
will enable the derivation of a 3D map of interstellar reddening.

\section{Flux calibration model} 
\label{pancino-sec-model}

Calibrating (spectro)photometry obtained from the usual type of ground based
observations (broadband imaging, spectroscopy) is not a trivial task, but  the
procedures are well known \citep[see, e.g.,][]{bessell99} and many groups have
developed sets of appropriate standard stars for the more than 200 photometric
known systems, and for spectroscopic observations.

In the case of Gaia, several instrumental effects -- much more complex than those
usually encountered -- redistribute light along the SED (Spectral Energy
Distribution) of the observed objects. The most difficult Gaia data to calibrate
are the BP and RP slitless spectra, requiring a new approach to the derivation of
the calibration model and to the SPSS grid needed to perform the actual
calibration. Some important complicating effects are:

\begin{itemize}
\item{the large focal plane with its large number of CCDs makes it so that
different observations of the same star will be generally on different CCDs, with
different quantum efficiencies, optical distorsions, transmissivity and so on.
Therefore, each wavelength and each position across the focal plane has its
(sometimes very different) PSF (point spread function);} 
\item{TDI (Time Delayed Integration) continuous reading mode, combined with the
need of compressing most of the data before on-ground transmission, make it
necessary to translate the full PSF into a linear (compressed into 1D) LSF (Line
Spread Function), which of course adds complication into the picture;}
\item{in-flight instrument monitoring is foreseen, but never comparable to the full
characterization that will be performed before launch, so the real instrument -- at
a  certain observation time -- will be slightly different from the theoretical one
assumed initially, and this difference will change with time;} 
\item{finally, radiation damage (or CTI, Charge Transfer Inefficiencies)
deserves special mention, for it is one of the most important factors in the
time variation of the instrument model \citep{weiler11,prod'homme11}.
It has particular impact onto the BP and RP dispersed images because the objects
travel along the BP and RP CCD strips in a direction that is parallel to the
spectral dispersion (wavelength coordinate) and therefore the net effect of
radiation damage can be to alter the SED of some spectra. Several solutions are
being tested to mitigate CTI effects, but the global instrument complexity
calls for a new approach to spectra flux calibrations.}
\end{itemize}

A flux calibration model is currently implemented in the photometric pipeline,
which splits the calibration into an {\em internal} and an {\em external} part.
The internal calibration model \citep{jordi11} uses a large number of well
behaved stars (internal standards), observed by Gaia, to report all observations
to a {\em reference} instrument, on the same instrumental relative flux and
wavelength scales. Once each observation for each object is reported to the
internal reference scales, the absolute or {\em external} calibration
\citep{PMN-003,SR-003} will use an appropriate SPSS set to report the relative
flux scale to an absolute flux scale in physical units, tied to the calibration
of Vega (see also Section~\ref{pancino_sec_spss}). Alternative approaches, where
the internal and external calibration steps are more inter-connected, are being
tested to maximise the precision and the accuracy of the Gaia calibration
\citep{AB-020}. The Gaia calibration model was also described by
\citet{pancino10}, \citet{jordi11}, and \citet{cacciari11}.

The final flux calibrated products will be: averaged (on all transits -- or
observations) white light magnitudes, G; integrated BP/RP magnitudes,
G$_{\rm{BP}}$ and G$_{\rm{RP}}$; flux calibrated BP/RP spectra; RVS spectra and
integrated G$_{\rm RVS}$ magnitudes, possibly also flux calibrated. The
G$_{\rm{BP}}$--G$_{\rm{RP}}$ colour will be used to correct for chromaticity
effects in the global astrometric solution. As said, only for specific classes of
objects, epoch spectra and magnitudes will be released, with variable stars as an
obvious example. 

The external calibration model contains -- as discussed -- a large number of
parameters, requiring a large number (about 200) of calibrators. With the
standard calibration techniques \citep{bessell99}, the best possible calibrators
are hot, almost featureless stars such as WD or hot subdwarfs. Unfortunately,
these stars are all similar to each other, forming an intrinsically degenerate
set. The Gaia calibration model instead requires to differentiate as much as
possible the calibrators, by including smooth spectra, but also spectra with
absorption features, both narrow (atomic lines) or wide (molecular bands),
appearing both on the blue and the red side of the spectrum\footnote{Including
emission line objects in our set of calibrators is problematic. Emission line
stars are often variable and thus do not make good calibrators. Similarly for
Quasars, which are typically faint for our ground-based campaigns. Thus, with
this calibration model we do not expect to be able to calibrate with very high
accuracy emission line objects.}. An experiment described by \citet{pancino10}
shows that the inclusion of just a few M stars\footnote{While M giants show
almost always variations of the order of 0.1--0.2~mag, and thus are not useful as
flux standards, M dwarfs rarely do \citep{eyer08}.} with large molecular
absorptions in the Gaia SPSS set can improve the calibration of similarly red
stars by a factor of more than ten (from a formal error of 0.15~mag to an error
smaller than 0.01~mag). 

\begin{figure}[!t]
\centering
\includegraphics[width=8cm]{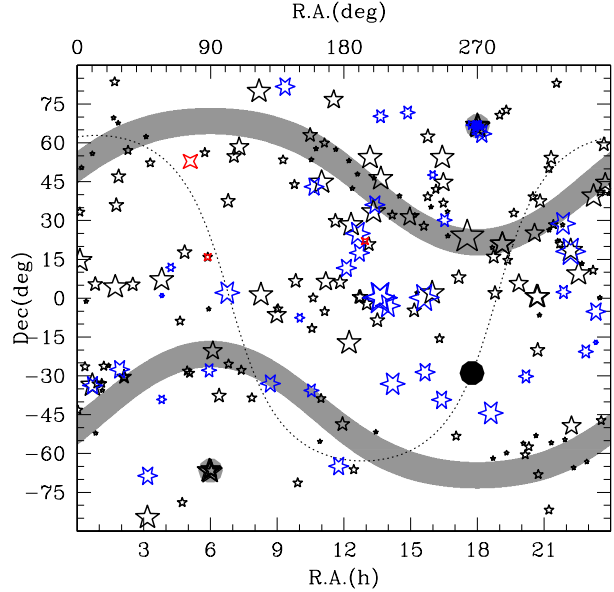}
\caption{Distribution of our SPSS candidates on the sky. The Galactic plane and
center are marked with a dotted line and a large black circle, respectively. The
Ecliptic poles are marked as two large grey circles, and two stripes at
$\pm$45~deg from the Ecliptic poles (roughly where Gaia is observing more often)
are shaded in grey. Our {\em Pillars} are shown as three four-pointed stars, the
{\em Primary SPSS candidates} as six-pointed stars, and the {\em Secondary SPSS
candidates} as five-pointed stars. The stars size is proportional to the SPSS
brightness, ranging from V$\simeq$8 (largest symbols) to 15~mag (smallest
symbols), approximately.}
\label{fig_sky}
\end{figure}

In conclusion, the complexity of the instrument reflects in a complex calibration
model, that requires a large set of homogeneously calibrated SPSS, covering a
range of spectral types. No such database exists in the literature, and new
observations are necessary to build it.

\section{The Gaia grid of spectrophotometric standard stars} 
\label{pancino_sec_spss}

From the above discussion, it is clear that the Gaia SPSS grid has to be chosen
with great care. The Gaia SPSS, or better their reference flux tables should
conform to the following general requirements (van Leeuven et al. 2010):

\begin{itemize}
\item{Resolution R=$\lambda/\delta\lambda \simeq 1000$, i.e., they should 
oversample the Gaia BP/RP resolution by a factor of  4--5 at least;}
\item{Wavelength coverage: 330--1050~nm;}
\item{Typical uncertainty on the absolute flux scale, with respect to the assumed
calibration of Vega, of a few percent, excluding small troubled areas in the
spectral range (telluric bands residuals, extreme red and blue edges), where it
can be somewhat worse.}
\end{itemize}

The total number of SPSS in the Gaia grid should be of the order of 200--300
stars, including a variety of spectral types. Clearly, no such large and
homogeneous dataset exists in the literature yet\footnote{The CALSPEC database
\citep{bohlin07} is not large enough for our purpose, especially considering the
strict criteria described below. Its extension to more than 100 SPSS is eagerly
awaited, but still not available to the public.}. It is therefore necessary to
build the Gaia SPSS grid with new, dedicated observations. We describe the
characteristics of the Gaia SPSS and of the dedicated observing campaigns in the
following Sections.

\subsection{SPSS Candidates}

We have followed a two steps approach \citep{MBZ-001} that firstly creates a set
of {\em Primary SPSS}, i.e., well known SPSS that are calibrated on the three
{\em Pillars} of the
CALSPEC\footnote{http://www.stsci.edu/hst/observatory/cdbs/calspec.html} set,
described in \citet{bohlin95} and \citet{bohlin07}, and tied to the Vega flux
calibration by \citet{bohlin04} and \citet{bohlin07}. The Primary SPSS
\citep{GA-001} will constitute the ground-based calibrators of the actual Gaia
grid, and need to be suitably bright for 2--4~m class telescopes in both
hemispheres. The most important sources for Primary candidates are the CALSPEC
grid, \citet{oke90}, \citet{hamuy92,hamuy94}, Stritzinger et al. (2005) and
others.

The Secondary SPSS \citep{GA-003} will form, together with the eligible
primaries, the actual Gaia SPSS grid and will conform to the following
requirements \citep{FVL-072}:

\begin{itemize}
\item{Secondary SPSS have spectra as featureless as possible (but see below
for exceptions);}
\item{Secondary SPSS shall be validated against variability;}
\item{The magnitude and sky location \citep[i.e., number of useful, clean
transits, see][]{JMC-001,JMC-002} of Secondary SPSS grants a resulting
S/N$\simeq$100 per sample over most of the wavelength range when observed by
Gaia (end of mission);}
\item{Secondary SPSS cover a range of spectral types and spectral shapes, as
needed to ensure the best possible claibration of all kinds of objects observed by
Gaia.}
\end{itemize}

Additional, special members of the Secondary SPSS candidates are: (1) a few
selected SPSS around the Ecliptic Poles, two regions of the sky  that will be
repeatedly observed by Gaia, in the first two weeks after reaching its orbit in
L2, for calibration purposes; (2) a few M stars with deep absorption features in
the red; (3) a few SDSS stars that have been observed in SEGUE sample
\citep{yanny09,MBZ-002}; (4) a few well known SPSS that are among the targets of
the ACCESS mission \citep{kaiser10}, dedicated to the absolute flux measurement
of a few stars besides Vega.

\section{The Gaia SPSS survey}
\label{pancino-sec-survey}

Our survey is split into two campaigns, the {\em main campaign} dedicated to
obtaining spectrophotometry of all our candidate SPSS, and the {\em auxiliary
campaign} dedicated to monitoring the constancy of our SPSS on relevant
timescales.

\begin{itemize}
\item{{\bf Main campaign.} Classical spectrophotometry \citep{bessell99} would
clearly be the best approach to obtain absolutely calibrated flux spectra if we
had a dedicated telescope. However tihs approach would require too much time,
given that we need high S/N of 300 stars, in photometric sky conditions, which
are rare. We thus decided for a combined approach, in which spectra are obtained
even if the sky is non-photometric\footnote{The cloud coverage must produce grey
extinction variations, i.e., the extinction must not alter significantly the
spectral shape. This condition is almost always verified in the case of veils or
thin clouds \citep{oke90,pakstiene03}, and can be checked a posteriori for each
observing night.}, providing the correct spectral shape of our SPSS. Then,
imaging in photometric conditions and in three bands (generally B, V, and R, but
sometimes also I and, more rarely, U) is obtained and calibrated magnitudes are
used to scale the spectra to the correct zeropoint.}
\item{{\bf Constancy monitoring.} Even stars used for years as spectrophotometric
standards were found to vary when dedicated studies have been performed 
\citep[see e.g., G24-9, that was found to be an eclipsing binary
by][]{landolt07}; our own survey has already found a few variables, including one
of the CALSPEC standards (Section~\ref{pancino-sec-res}). White dwarfs may show
variability with (multi-)periods from about 1 to 20~min and amplitudes from 
about 1-2\% up to 30\%, i.e., ZZ~Ceti type variability. We have tried to exclude
stars within the instability strips for DAV \citep{castanheira07}, DBV, and DOV
but in many cases the existing information was not sufficient (or sufficiently
accurate) to firmly establish the constant nature of a given WD. Also redder
stars are often found to be variable: for example K stars have shown variability
of 5-10\% with periods of the order 1--10 days \citep{eyer97}. In addition,
eclipsing binaries are frequent and can be found at all spectral types. Their
periods can range from a few hours to hundreds of days, most of them
having $P\simeq$ 1-10~days, \citep{dvorak04}. } 
\end{itemize}

Our chosen observing facilities are imagers and low-resolution spectrographs in
both hemispheres: EFOSC2@NTT and ROSS@REM at the ESO La Silla Observatory, Chile;
CAFOS@2.2m at the Calar Alto Observatory, Spain; DOLORES@TNG at the Roque de Los
Muchachos in La Palma, Spain; LaRuca@1.5m at the San Pedro M\'artir Observatory,
Mexico; BFOSC@Cassini in Loiano, Italy. Observations are almost complete --- at
the time of writing --- only the long-term (3~yr) monitoring will be ongoing
after Summer 2012, and is expected to finish in 2013-2014.

\begin{figure}
\centering
\includegraphics[width=10cm]{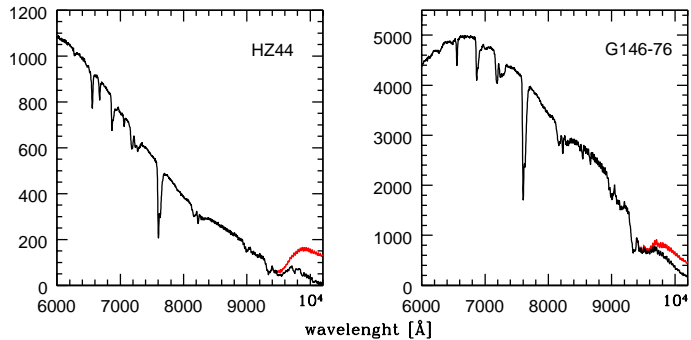}
\caption{Second-order contamination on DOLoRes@TNG spectra of a blue star (left
panel) and a red star (right panel); the black lines are the corrected spectra,
while the red lines above, starting at about 9500~\AA, show the contaminated
spectra.}
\label{pancino_fig_2nd}
\end{figure}

\section{Data treatment and data products}
\label{sec-reds}

The required precision and accuracy of the SPSS calibration imposes the adoption
of strict protocols of instrument characterization, data reduction, quality
control, and data analysis. In particular, we carried out a detailed
characterization of the intruments used:

\begin{itemize}
\item{CCD familiarization plan, containing a study of the dark and bias frames
stability; the shutter characterization (shutter times and delays); and the study
of the linearity of all employed CCDs;}
\item{Instrument familiarization plan studying the stability of imaging and
spectroscopy flats, the study of fringing, and the lamp flexures of the employed
spectrographs;}
\item{Site familiarization plan, providing extinction curves, extinction
coefficients, colour terms, and a study of the effect of
``calima"\footnote{Calima is a dust wind originating in the Sahara air layer,
which often affects observations in the Canary Islands.} on the spectral shape.}
\end{itemize}

As a results of these studies, specific recommendations for observations and
data treatment were defined. Data reductions are being performed mostly with
IRAF\footnote{IRAF is the Image Reduction and Analysis Facility, a general
purpose software system for the reduction and analysis of astronomical data.
IRAF is written and supported by the IRAF programming group at the National
Optical Astronomy Observatories (NOAO) in Tucson, Arizona. NOAO is operated by
the Association of Universities for Research in Astronomy (AURA), Inc. under
cooperative agreement with the National Science Foundation} and IRAF-based
pipelines. We also use SExtractor \citep{bertin96} for aperture photometry,
because it provides many useful parameters that we will use for a semi-automated
quality control (QC) of each reduced frame.

To illustrate a detail of the reduction procedures, we show in
Figure~\ref{pancino_fig_2nd} our second-order contamination correction for a
blue and a red star. The effect arises when light from blue wavelengths, from
the second dispersed order of a particular grism or grating, falls on the red
wavelengths of the first dispersed order. Such contamination usually happens
when the instrument has no cross-disperser. Of the instruments we use, only
EFOSC2@NTT and DOLoRes@TNG present significant contamination. To map the blue
light falling onto our red spectra, we adapted a method proposed by
\citet{sanchez06} and applied it to dedicated observations. Our wavelength maps
generally allow to recover the correct spectral shape with residuals within
$\pm$2\%, as tested on a few CALSPEC standards observed with both TNG and NTT. 

\begin{figure}
\centering
\includegraphics[width=8cm]{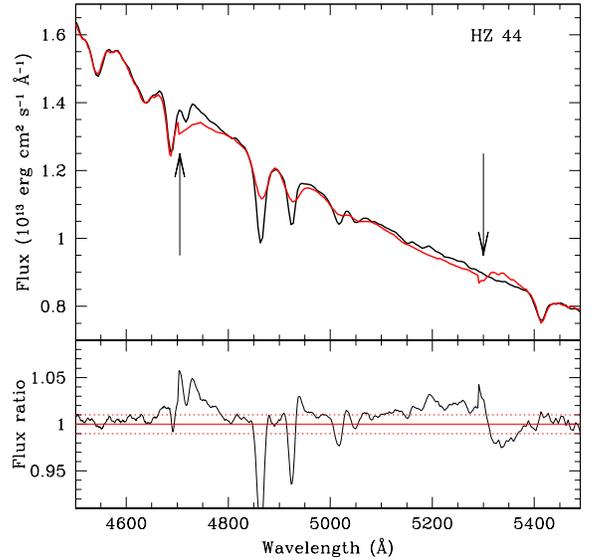}
\caption{Top panel: comparison of our preliminary spectrum of HZ~44 (thick black
line) with the CALSPEC tabulated spectrum (thick red line) in a region where we
found a discrepancy (marked by the two arrows), where small $\simeq$0.5--1.0\%
jumps in the CALSPEC spectrum are probably due to a mismatch of two different
spectra. Bottom panel: ratio between our spectrum and the CALSPEC spectrum;
perfect agreement (red line) and $\pm$1\% agreement (dotted red lines, our
requirement) are marked.}
\label{pancino_fig_hz44}
\end{figure}

As an example of the final data quality, we show in
Figure~\ref{pancino_fig_hz44} a test performed to refine our reduction
procedures, where a portion of the spectrum of HZ44 observed in a photometric
night is compared with the CALSPEC flux table. We point out that this was just a
preliminary reduction, where fringing and telluric absorption features were not
properly removed yet. Even with these limitations, we were able to meet the
requirements, because the residuals between our spectrum and the CALSPEC
tabulated one were on average lower than 1\%, with the exception of the low S/N
red edge and of the telluric absorption bands. However, some unsatisfactory
jumps appeared in the comparison, between 4000 and 6000~\AA, where our spectra
have the highest S/N. As shown in Figure~\ref{pancino_fig_hz44}  (top panel) and
already noted by \citet{bohlin01}, the jumps were due to a (minor) problem in
the CALSPEC spectrum, probably where two pieces of the spectrum were joined. 

\subsection{Preliminary results}
\label{pancino-sec-res}

During the years, we have cleaned our initial linelists \citep{GA-001,GA-003}
from unsuitable stars. On the one hand we kept our literature information
updated, and to look for new about variability, activity, close companions and
misidentification. On the other hand we used our own data to test the validity
of each candidate. 

Identification problems are common, especially when large databases are
automatically matched (as done within SIMBAD, for example), and when stars have
large proper motions. We found a few wrong identifications, the most notable case
being that of LTT~377, which showed an F type spectrum rather than the expected
K, and was in many databases confused with star CD~-34~241\footnote{At the moment
of writing, the SIMBAD database has been updated and now the correct
identification is reported.}, which corresponds to CD~-34~239. A similar case was
WD~0204-306 for which we obtained an unexpectedly red spectrum, corresponding to
that of star LP~885-23, uncorrectly identified with WD~0204-306, while it is
instead LP~885-22. A more critical example was WD~1148-230, having very different
coordinates in the \citet{mccook99} catalogue \citep[coming from][]{stys00} and
in SIMBAD. The SIMBAD coordinates were from the 2MASS catalogue \citep{cutri03}.
Magnitudes were also significantly different, so we had to reject the candidate.

In a few cases candidates that appeared relatively isolated on the available
finding charts turned out to be in a crowded area where no aperture photometry
or reliable wide slit spectroscopy could be performed from the ground, or showed
previously unseen companions. Generally, stars with high proper motion could
appear isolated in some past finding chart, but later moved too close to another
star to be safely observed from the ground. Examples of candidates showing the
presence of previously unknown and relatively bright companions were WD~0406+592
and WD~2058+181. Both stars were rejected.

\begin{figure}
\centering
\includegraphics[angle=270,width=10cm]{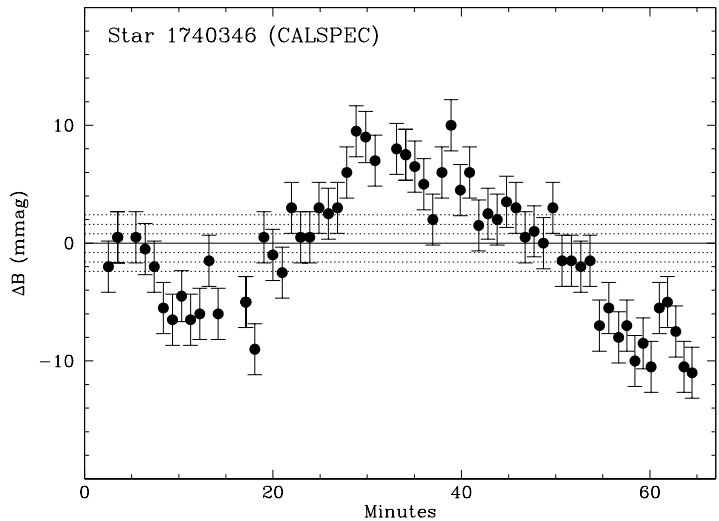}
\caption{Our best lightcurve for the CALSPEC standard 1740346 (obtained with
BFOSC in Loiano on 1 September 2010), originally one of our {\em Primary SPSS}
candidates. The average of all field-stars magnitude differences (i.e., zero) is
marked with a solid line, while the $\pm$1, 2, and 3~$\sigma$ variations are
marked with dotted lines.}
\label{pancino_fig_var}
\end{figure}

Finally, star 1740346, one of the currently used CALSPEC standards and one of our
{\em Primary SPSS} candidates, showed variability with an amplitude of
10~$\pm$~0.8~mmag in B band when observed with BFOSC@Cassini in Loiano, on 1
September 2010; with DOLoRes@TNG, on 31 September 2009; and with BFOSC@Cassini,
on 26 May 2009. The variability period is 50~min, approximately. A preliminary
determination of 1740346 parameters can be found in \citet{marinoni11}, using
literature data and stellar models, resulting in a mass of
$\simeq$1.3~M$_{\odot}$, an effective temperature of $\simeq$8300~K, and a
distance of $\simeq$750~pc. These parameters are also compatible with a
$\delta$~Scuti type star. We are gathering detailed follow-up observations for a
complete characterization of star 1740346. Our best differential lightcurve is
presented in Figure~\ref{pancino_fig_var}.

\section{Summary and conclusions}

The Gaia mission and its data reduction is a challenging enterprise, carried on
by ESA and the European scientific community. An even more challenging enterprise
will be the scientific interpretation of such a large database, and the creation
of tools which can efficiently explore it. 

As an example of the DPAC (Data Processing and Analysis Consortium) tasks, I have
briefly described the Gaia flux calibration model and the Gaia SPSS survey, which
will build a large ($\simeq200-300$) grid of SPSS with 1--3\% flux calibration
with respect to Vega. All data products will be eventually made public together
with each Gaia data release, within the framework of the DPAC publication
policies. At the moment the accumulated data and literature information are
stored locally and can be accessed upon request.


\bibliography{Pancino_E_2}

\begin{thebibliography}{}
\expandafter\ifx\csname natexlab\endcsname\relax\def\natexlab#1{#1}\fi
\expandafter\ifx\csname url\endcsname\relax
  \def\url#1{\texttt{#1}}\fi
\expandafter\ifx\csname urlprefix\endcsname\relax\def\urlprefix{URL }\fi
\providecommand{\eprint}[2][]{\url{#2}}

\bibitem[{{Altavilla} et~al.(2008){Altavilla}, {Bellazzini}, {Pancino},
  {Bragaglia}, {Cacciari}, {Diolaiti}, {Federici}, {Montegriffo}, \&
  {Rossetti}}]{GA-001}
{Altavilla}, G., {Bellazzini}, M., {Pancino}, E., {Bragaglia}, A., {Cacciari},
  C., {Diolaiti}, E., {Federici}, L., {Montegriffo}, P., \& {Rossetti}, E.
  2008, {Primary standards for the establishment of the Gaia grid of SPSS.
  Selection criteria and list of candidates.}, Tech. Rep.
  GAIA-C5-TN-OABO-GA-001

\bibitem[{{Altavilla} et~al.(2010){Altavilla}, {Bragaglia}, {Pancino},
  {Bellazzini}, {Cacciari}, {Federici}, \& {Ragaini}}]{GA-003}
{Altavilla}, G., {Bragaglia}, A., {Pancino}, E., {Bellazzini}, M., {Cacciari},
  C., {Federici}, L., \& {Ragaini}, S. 2010, {Secondary standards for the
  establishment of the Gaia grid of SPSS. Selection criteria and list of
  candidates.}, Tech. Rep. GAIA-C5-TN-OABO-GA-003

\bibitem[{{Bellazzini} et~al.(2010){Bellazzini}, {Altavilla}, \&
  {Cacciari}}]{MBZ-002}
{Bellazzini}, M., {Altavilla}, G., \& {Cacciari}, C. 2010, {Notes on the
  possible use of SEGUE spectrophotometry for the absolute photometric
  calibration of Gaia.}, Tech. Rep. GAIA-C5-TN-OABO-MBZ-002

\bibitem[{{Bellazzini} et~al.(2006){Bellazzini}, {Bragaglia}, {Federici},
  {Diolaiti}, {Cacciari}, \& {Pancino}}]{MBZ-001}
{Bellazzini}, M., {Bragaglia}, A., {Federici}, L., {Diolaiti}, E., {Cacciari},
  C., \& {Pancino}, E. 2006, {Absolute calibration of Gaia photometric data. I.
  General considerations and requirements.}, Tech. Rep. GAIA-C5-TN-OABO-MBZ-001

\bibitem[{{Bertin} \& {Arnouts}(1996)}]{bertin96}
{Bertin}, E., \& {Arnouts}, S. 1996, \aaps, 117, 393

\bibitem[{{Bessell}(1999)}]{bessell99}
{Bessell}, M.~S. 1999, \pasp, 111, 1426

\bibitem[{{Bohlin}(2007)}]{bohlin07}
{Bohlin}, R.~C. 2007, in The Future of Photometric, Spectrophotometric and
  Polarimetric Standardization, edited by C.~{Sterken}, vol. 364 of
  Astronomical Society of the Pacific Conference Series, 315.
  \eprint{arXiv:astro-ph/0608715}

\bibitem[{{Bohlin} et~al.(1995){Bohlin}, {Colina}, \& {Finley}}]{bohlin95}
{Bohlin}, R.~C., {Colina}, L., \& {Finley}, D.~S. 1995, \aj, 110, 1316

\bibitem[{{Bohlin} et~al.(2001){Bohlin}, {Dickinson}, \& {Calzetti}}]{bohlin01}
{Bohlin}, R.~C., {Dickinson}, M.~E., \& {Calzetti}, D. 2001, \aj, 122, 2118

\bibitem[{{Bohlin} \& {Gilliland}(2004)}]{bohlin04}
{Bohlin}, R.~C., \& {Gilliland}, R.~L. 2004, \aj, 127, 3508

\bibitem[{{Brown} et~al.(2010){Brown}, {Jordi}, \& {Fabricius}}]{AB-020}
{Brown}, A., {Jordi}, C., \& {Fabricius}, C. 2010, {Forward modeling of the
  BP/RP data processing: options and implications.}, Tech. Rep.
  GAIA-C5-TN-LEI-AB-020

\bibitem[{{Cacciari}(2011)}]{cacciari11}
{Cacciari}, C. 2011, in EAS Publications Series, vol.~45 of EAS Publications
  Series, 155

\bibitem[{{Carrasco} et~al.(2010){Carrasco}, {Jordi}, {Figueras}, \&
  B.}]{JMC-001}
{Carrasco}, J.~M., {Jordi}, C., {Figueras}, A.~G., F., \& B., A.~E. 2010,
  {Towards the selection of standard stars for absolute flux calibration.
  Signal-to-noise ratios for BP/RP spectra and crowding due to FoV
  overlapping.}, Tech. Rep. GAIA-C5-TN-UB-JMC-001

\bibitem[{{Carrasco} et~al.(2007){Carrasco}, {Jordi}, {Lopez-Marti},
  {Figueras}, \& {Anglada-Escude}}]{JMC-002}
{Carrasco}, J.~M., {Jordi}, C., {Lopez-Marti}, B., {Figueras}, F., \&
  {Anglada-Escude}, G. 2007, {Revolving phase effct to FoV overlapping and its
  application to primary SPSS.}, Tech. Rep. GAIA-C5-TN-UB-JMC-002

\bibitem[{{Castanheira} et~al.(2007){Castanheira}, {Kepler}, {Costa},
  {Giovannini}, {Robinson}, {Winget}, {Kleinman}, {Nitta}, {Eisenstein},
  {Koester}, \& {Santos}}]{castanheira07}
{Castanheira}, B.~G., {Kepler}, S.~O., {Costa}, A.~F.~M., {Giovannini}, O.,
  {Robinson}, E.~L., {Winget}, D.~E., {Kleinman}, S.~J., {Nitta}, A.,
  {Eisenstein}, D., {Koester}, D., \& {Santos}, M.~G. 2007, \aap, 462, 989.
  \eprint{arXiv:astro-ph/0611332}

\bibitem[{{Cutri} et~al.(2003){Cutri}, {Skrutskie}, {van Dyk}, {Beichman},
  {Carpenter}, {Chester}, {Cambresy}, {Evans}, {Fowler}, {Gizis}
  et~al.}]{cutri03}
{Cutri}, R.~M., {Skrutskie}, M.~F., {van Dyk}, S., {Beichman}, C.~A.,
  {Carpenter}, J.~M., {Chester}, T., {Cambresy}, L., {Evans}, T., {Fowler}, J.,
  {Gizis}, J., et~al. 2003, VizieR Online Data Catalog, 2246, 0

\bibitem[{{Dvorak}(2004)}]{dvorak04}
{Dvorak}, S.~W. 2004, Information Bulletin on Variable Stars, 5542, 1

\bibitem[{{Eyer} \& {Grenon}(1997)}]{eyer97}
{Eyer}, L., \& {Grenon}, M. 1997, in Hipparcos - Venice '97, edited by R.~M.
  {Bonnet}, E.~{H{\o}g}, P.~L. {Bernacca}, L.~{Emiliani}, A.~{Blaauw},
  C.~{Turon}, J.~{Kovalevsky}, L.~{Lindegren}, H.~{Hassan}, M.~{Bouffard},
  B.~{Strim}, D.~{Heger}, M.~A.~C. {Perryman}, \& L.~{Woltjer}, vol. 402 of ESA
  Special Publication, 467

\bibitem[{{Eyer} \& {Mowlavi}(2008)}]{eyer08}
{Eyer}, L., \& {Mowlavi}, N. 2008, Journal of Physics Conference Series, 118,
  012010. \eprint{0712.3797}

\bibitem[{{Hamuy} et~al.(1994){Hamuy}, {Suntzeff}, {Heathcote}, {Walker},
  {Gigoux}, \& {Phillips}}]{hamuy94}
{Hamuy}, M., {Suntzeff}, N.~B., {Heathcote}, S.~R., {Walker}, A.~R., {Gigoux},
  P., \& {Phillips}, M.~M. 1994, \pasp, 106, 566

\bibitem[{{Hamuy} et~al.(1992){Hamuy}, {Walker}, {Suntzeff}, {Gigoux},
  {Heathcote}, \& {Phillips}}]{hamuy92}
{Hamuy}, M., {Walker}, A.~R., {Suntzeff}, N.~B., {Gigoux}, P., {Heathcote},
  S.~R., \& {Phillips}, M.~M. 1992, \pasp, 104, 533

\bibitem[{{Ibata} \& {Gibson}(2007)}]{ibata07}
{Ibata}, R., \& {Gibson}, B. 2007, Scientific American, 296, 040000

\bibitem[{{Jordi}(2011)}]{jordi11}
{Jordi}, C. 2011, in EAS Publications Series, vol.~45 of EAS Publications
  Series, 149

\bibitem[{{Kaiser} et~al.(2010){Kaiser}, {Kruk}, {McCandliss}, {Rauscher},
  {Kimble}, {Pelton}, {Sahnow}, {Dixon}, {Feldman}, {Gaither}
  et~al.}]{kaiser10}
{Kaiser}, M.~E., {Kruk}, J.~W., {McCandliss}, S.~R., {Rauscher}, B.~J.,
  {Kimble}, R.~A., {Pelton}, R.~S., {Sahnow}, D.~J., {Dixon}, W.~V., {Feldman},
  P.~D., {Gaither}, et~al. 2010, in Society of Photo-Optical Instrumentation
  Engineers (SPIE) Conference Series, vol. 7731 of Society of Photo-Optical
  Instrumentation Engineers (SPIE) Conference Series

\bibitem[{{Landolt} \& {Uomoto}(2007)}]{landolt07}
{Landolt}, A.~U., \& {Uomoto}, A.~K. 2007, \aj, 133, 768. \eprint{0704.3030}

\bibitem[{{Marinoni}(2011)}]{marinoni11}
{Marinoni}, S. 2011, Ph.D. thesis, INAF - Osservatorio Astronomico di Roma +
  ASI Science Data Center <EMAIL>silvia.marinoni@asdc.asi.it</EMAIL>

\bibitem[{{McCook} \& {Sion}(1999)}]{mccook99}
{McCook}, G.~P., \& {Sion}, E.~M. 1999, VizieR Online Data Catalog, 3210, 0

\bibitem[{{Mignard}(2005)}]{mignard05}
{Mignard}, F. 2005, in Astrometry in the Age of the Next Generation of Large
  Telescopes, edited by P.~K. {Seidelmann}, \& A.~K.~B. {Monet}, vol. 338 of
  Astronomical Society of the Pacific Conference Series, 15

\bibitem[{{Montegriffo} \& {Bellazzini}(2009)}]{PMN-003}
{Montegriffo}, P., \& {Bellazzini}, M. 2009, {A model for the absolute
  photometric calibration of Gaia BP and RP spectra. III. A full in-flight
  calibration of the model parameters.}, Tech. Rep. GAIA-C5-TN-OABO-PMN-003

\bibitem[{{Oke}(1990)}]{oke90}
{Oke}, J.~B. 1990, \aj, 99, 1621

\bibitem[{{Pak{\v s}tiene} \& {Solheim}(2003)}]{pakstiene03}
{Pak{\v s}tiene}, E., \& {Solheim}, J.-E. 2003, Baltic Astronomy, 12, 221

\bibitem[{{Pancino}(2010)}]{pancino10}
{Pancino}, E. 2010, in Hubble after SM4. Preparing JWST. \eprint{1009.1748}

\bibitem[{{Perryman} et~al.(1997){Perryman}, {Lindegren}, \&
  {Turon}}]{perryman97}
{Perryman}, M.~A.~C., {Lindegren}, L., \& {Turon}, C. 1997, in Hipparcos -
  Venice '97, edited by R.~M. {Bonnet}, E.~{H{\o}g}, P.~L. {Bernacca},
  L.~{Emiliani}, A.~{Blaauw}, C.~{Turon}, J.~{Kovalevsky}, L.~{Lindegren},
  H.~{Hassan}, M.~{Bouffard}, B.~{Strim}, D.~{Heger}, M.~A.~C. {Perryman}, \&
  L.~{Woltjer}, vol. 402 of ESA Special Publication, 743

\bibitem[{{Prod'Homme}(2011)}]{prod'homme11}
{Prod'Homme}, T. 2011, in EAS Publications Series, vol.~45 of EAS Publications
  Series, 55

\bibitem[{{Ragaini} et~al.(2011){Ragaini}, {Montegriffo}, \&
  {Cacciari}}]{SR-003}
{Ragaini}, S., {Montegriffo}, P., \& {Cacciari}, C. 2011, {A new model for the
  absolute photometric calibration of AF integrated photometry.}, Tech. Rep.
  GAIA-C5-TN-OABO-SR-003

\bibitem[{{S{\'a}nchez-Bl{\'a}zquez} et~al.(2006){S{\'a}nchez-Bl{\'a}zquez},
  {Peletier}, {Jim{\'e}nez-Vicente}, {Cardiel}, {Cenarro},
  {Falc{\'o}n-Barroso}, {Gorgas}, {Selam}, \& {Vazdekis}}]{sanchez06}
{S{\'a}nchez-Bl{\'a}zquez}, P., {Peletier}, R.~F., {Jim{\'e}nez-Vicente}, J.,
  {Cardiel}, N., {Cenarro}, A.~J., {Falc{\'o}n-Barroso}, J., {Gorgas}, J.,
  {Selam}, S., \& {Vazdekis}, A. 2006, \mnras, 371, 703.
  \eprint{arXiv:astro-ph/0607009}

\bibitem[{{Stys} et~al.(2000){Stys}, {Slevinsky}, {Sion}, {Saffer}, {Holberg},
  {O'Donoghue}, {Kilkenny}, {Stobie}, \& {Koen}}]{stys00}
{Stys}, D., {Slevinsky}, R., {Sion}, E.~M., {Saffer}, R., {Holberg}, J.~B.,
  {O'Donoghue}, D., {Kilkenny}, D., {Stobie}, R.~S., \& {Koen}, C. 2000, \pasp,
  112, 354

\bibitem[{{van Leeuwen} et~al.(2011){van Leeuwen}, {Pancino}, \&
  {Altavilla}}]{FVL-072}
{van Leeuwen}, F., {Pancino}, E., \& {Altavilla}, G. 2011, {GB\_Obs software
  requirement specifications.}, Tech. Rep. GAIA-C5-SP-IOA-FVL-072

\bibitem[{{Weiler} et~al.(2011){Weiler}, {Babusiaux}, \& {Short}}]{weiler11}
{Weiler}, M., {Babusiaux}, C., \& {Short}, A. 2011, in EAS Publications Series,
  vol.~45 of EAS Publications Series, 67

\bibitem[{{Yanny} et~al.(2009){Yanny}, {Rockosi}, {Newberg}, {Knapp},
  {Adelman-McCarthy}, {Alcorn}, {Allam}, {Allende Prieto}, {An}, {Anderson}
  et~al.}]{yanny09}
{Yanny}, B., {Rockosi}, C., {Newberg}, H.~J., {Knapp}, G.~R.,
  {Adelman-McCarthy}, J.~K., {Alcorn}, B., {Allam}, S., {Allende Prieto}, C.,
  {An}, D., {Anderson}, et~al. 2009, \aj, 137, 4377. \eprint{0902.1781}

\end{thebibliography}

\end{document}